\documentclass{aastex63}

\received{}
\revised{}
\accepted{}

\submitjournal{ApJL}

\shorttitle{Scaling Laws Through the Inner Heliosphere}
\shortauthors{Alberti et al.}

\begin{document}

\title{On the Scaling Properties of Magnetic Field Fluctuations Through the Inner Heliosphere}

\correspondingauthor{Tommaso Alberti}
\email{tommaso.alberti@inaf.it}

\author[0000-0001-6096-0220]{Tommaso Alberti}
\affiliation{INAF - Istituto di Astrofisica e Planetologia Spaziali, via del Fosso del Cavaliere 100, 00133, Roma, Italy}

\author[0000-0001-5481-4534]{Monica Laurenza}
\affiliation{INAF - Istituto di Astrofisica e Planetologia Spaziali, via del Fosso del Cavaliere 100, 00133, Roma, Italy}

\author[0000-0002-3403-647X]{Giuseppe Consolini}
\affiliation{INAF - Istituto di Astrofisica e Planetologia Spaziali, via del Fosso del Cavaliere 100, 00133, Roma, Italy}

\author[0000-0002-0266-2556]{Anna Milillo}
\affiliation{INAF - Istituto di Astrofisica e Planetologia Spaziali, via del Fosso del Cavaliere 100, 00133, Roma, Italy}

\author[0000-0002-5002-6060]{Maria Federica Marcucci}
\affiliation{INAF - Istituto di Astrofisica e Planetologia Spaziali, via del Fosso del Cavaliere 100, 00133, Roma, Italy}

\author[0000-0002-3182-6679]{Vincenzo Carbone}
\affiliation{Universit\`a della Calabria, Dip. di Fisica, Ponte P. Bucci, Cubo 31C, 87036, Rende (CS), Italy}

\author[0000-0002-1989-3596]{Stuart D. Bale}
\affiliation{Space Sciences Laboratory, University of California, Berkeley, CA 94720-7450, USA}
\affiliation{Physics Department, University of California, Berkeley, CA 94720-7300, USA}

\begin{abstract}
Although the interplanetary magnetic field variability has been extensively investigated {\it in situ} by means of data coming from several space missions, the newly launched missions providing high-resolution measures and approaching the Sun, offer the possibility to study the multiscale variability in the innermost solar system. Here by means of the Parker Solar Probe measurements we investigate the scaling properties of solar wind magnetic field 
fluctuations at different heliocentric distances. The results show a clear transition at distances close to say $0.4$ au. Closer to the Sun fluctuations show a $f^{-3/2}$ frequency power spectra and regular scaling properties, while for distances larger than $0.4$ au fluctuations show a Kolmogorov spectrum $f^{-5/3}$ and are characterized by anomalous scalings. The observed statistical properties of turbulence suggests that the solar wind magnetic fluctuations, in the late stage far form the Sun, show a multifractal behaviour typical of turbulence and described through intermittency, while in the early stage, when leaving the solar corona, a breakdown of these properties are observed, thus showing a statistical monofractal global self-similarity. Physically the breakdown observed close to the Sun should be due either to a turbulence with regular statistics or to the presence of intense stochastic fluctuations able to cancel out correlations necessary for the presence of anomalous scaling. 
\end{abstract}


\section{Introduction} \label{sec:intro}

Since the 70s several space missions have been launched to provide new insights into the solar phenomena and solar wind properties (e.g., Helios, Ulysses, Wind, ACE) allowing us to collect a wide amount of data about the processes that cause the solar wind formation and evolution throughout the interplanetary space \citep[e.g.,][]{Rosenbauer77,Denskat82,Grappin90}. Among other topics \citep[e.g.,][]{Burlaga82,McComas95,Marsch18}, a wide attention has been paid to turbulence in the solar wind by investigating the scaling behavior of both velocity and magnetic field components \citep[e.g.,][and references therein]{Dobrowolny80,Matthaeus82,Tu90,Bruno13,Alberti19a}. Indeed, solar wind magnetic field fluctuations around the large-scale mean field, usually described within the magnetohydrodynamic (MHD) framework, are characterized by scale-invariant features over a wide range of scales \citep[e.g.,][]{Bruno13}. At 1 au, this range of scales, known as inertial range \citep{Kolmogorov41,Frisch95}, is dominated by Alfv\'enic fluctuations \citep{Belcher71,Bruno13} mixed with slow mode compressive ones \citep{Howes12,Klein12,Verscharen17}. This type of turbulence is characterized by an anisotropic cascade \citep{Horbury08,Chen16}, mostly described by models of balance and imbalanced Alfv\'enic turbulence \citep{Lithwick07,Perez09,Chandran15,Mallet17}, although different scalings are observed depending on several features as the role of the large-scale forcing \citep{Velli89}, the (im)balance between Alfv\'enic fluctuations \citep{Boldyrev06,Chandran15,Mallet17}, and so on \citep{Chen16,Chen20}. Moving closer to the Sun, a decreasing scaling slope is observed with a transition mostly occurring near 0.4 au \citep{Dobrowolny80,Denskat82,Tu90,Chen20}, the inertial range tends to move towards a more steady state \citep{Chen20}, an increase in the scale-dependent alignment and cross-helicity is also observed \citep{Boldyrev06,Lithwick07}, together with a different nature of the nonlinear coupling between different frequencies and/or damping/propagation effects \citep[e.g.,][]{Dobrowolny80}. Moreover, as the Sun is approached an increase of up to two order of magnitude is observed for turbulence energy, together with less steep spectra for magnetic field components, the velocity field and the Els\"asser variables, being characterized by a spectral exponent closer to -3/2 \citep{Chen20}. Furthermore, the role of slow-mode fluctuations tend to be reduced as for the rate of compressible magnetic fluctuations, while outward-propagating Alfv\'enic perturbations dominate on inward-propagating ones, consistent with turbulence-driven models \citep{Boldyrev06,Chandran15,Mallet17}.

Nowadays, a large amount of spacecraft, providing more accurate in situ measurements through high-resolution instruments, is available for monitoring the evolution of solar wind parameters and for providing new insights into the physics of the Sun and the solar wind. Furthermore, the different locations and orbits of these spacecraft could offer the possibility of investigating some interesting properties of solar wind turbulence and its evolution throughout the heliosphere \citep[e.g.,][]{Nicolaou19}, especially going as near as possible to the solar surface \citep{Marsden03,Fox16}. The recently launched missions, e.g., {\it Parker Solar Probe} (PSP), {\it BepiColombo}, and {\it Solar Orbiter}, and the in situ orbiting ones, e.g., ACE, Wind, and STEREO, offer the unique opportunity of multi-spacecraft combined observations of the interplanetary medium variability, the evolution of turbulence and solar wind structures at different distances from the Sun, the interaction between the solar wind plasma and planetary environments, and so on \citep[e.g.,][]{Milillo10,Muller13,Howard19,Kasper19,McComas19}. Recently, in the framework of solar wind turbulence \citet{Chen20} investigated the behavior of the power spectral density at different heliocentric distances by means of the first two orbits of the Parker Solar Probe spacecraft showing that the power-law spectral index moves from $\alpha_B \sim$ -3/2 to $\alpha_B \sim$ -5/3 when passing from $r \sim$ 0.17 au to $r \sim$ 0.6 au.

In this manuscript we deal with the analysis of the interplanetary magnetic field fluctuations along the PSP trajectory during its first and second orbits towards the Sun by means of a novel formalism based on the Hilbert Spectral Analysis (HSA). Specifically, we investigate the $q-$order scaling features of magnetic field components at different heliocentric distances (Section \ref{sec:methods}). In Section \ref{sec:results}, the results show that the inertial range scaling properties significantly change when moving from closer to farther the Sun, with intermittency completely emerging at distances larger than 0.4 au. Indeed, scaling exponents show a linear behavior at smaller heliocentric distances, while larger exponents, being characterized by a nonlinear convex behavior with the statistical order $q$, are found at $r >$ 0.4 au. In Section \ref{sec:conclusions}, we conclude that the result of this study could open new perspectives for describing the fractal properties of solar wind and to correctly characterize turbulence and intermittency in space plasmas at different locations.

\section{Data} \label{sec:data}

For this study we use solar wind magnetic field components in the heliocentric RTN reference frame (R=radial, T=tangential, N=normal) as measured by the PSP magnetometer. The PSP magnetic field data are taken by the outboard FIELDS Fluxgate Magnetometer (MAG) \citep{Bale16,Bale19} and are averaged to 1-s cadence from their native 4 samples per cycle cadence \citep{Fox16}. Data were freely retrieved from the Space Physics Data Facility (SPDF) Coordinated Data Analysis Web (CDAWeb) interface at \url{https://cdaweb.gsfc.nasa.gov/index.html/}. 

\begin{figure}[ht!]
\plotone{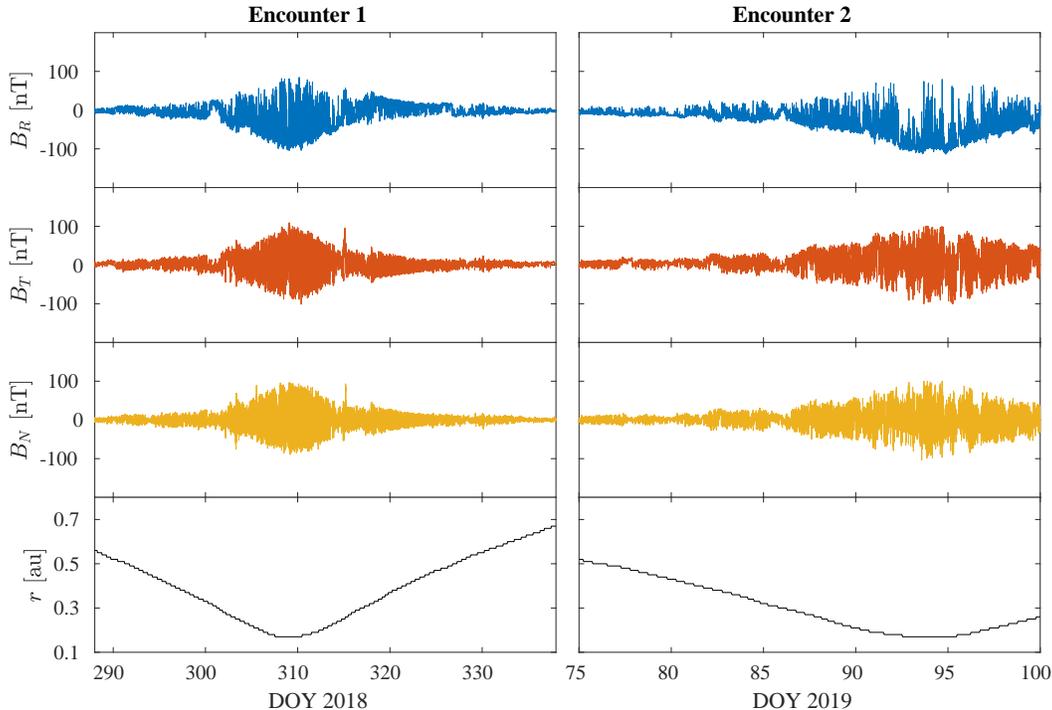}
\caption{(From top to bottom) The three components of the interplanetary magnetic field (at 1-s resolution), and (lower panel) the PSP radial distance from the Sun (at 1-hr resolution). The blue, orange, and yellow lines refer to the radial, tangential, and normal components, respectively. The right and the left panels show measurements during the first and the second PSP orbits approaching the Sun, respectively. \label{fig1}}
\end{figure}

For investigating the evolution of the interplanetary magnetic field we used the first and the second orbit of PSP towards the Sun, only considering adjacent temporal measurements during which no data gaps were found (i.e., the best time coverage of the FIELDS instrument). These corresponds to the period between 15 October and 04 December, 2018, and between 16 March and 10 April, 2019, for the first and the second orbits, respectively. During the intervals of investigation the solar wind speed was between 250 km/s and 650 km/s and the proton density ranged between $n\sim10$ cm$^{-3}$ (at 0.7 au) and $\sim400$ cm$^{-3}$ (at 0.17 au). Figure \ref{fig1} shows the three components of the interplanetary magnetic field (at 1-s resolution) and the PSP radial distance from the Sun (at 1-hr resolution). 

It is clear that magnetic field fluctuations decrease with increasing heliocentric distance of about one order of magnitude \citep[i.e., $B(r) \sim 1/r^2$,][]{Parker58}. However, by simply looking at the time series it is not sufficient to clearly discriminate between the different dynamical regimes and their evolution at different heliocentric distances, that is a crucial point for correctly characterizing dynamical processes such as the evolution of turbulence and intermittency, the large-scale structures dynamics, the mean field approximation, and so on.  

\section{Methods} \label{sec:methods}

Investigating field fluctuations is usually one of the most important aspects of dealing with the existence of dynamical processes and phenomena characterizing physical systems. Generally, this can be achieved by means of data analysis methods allowing us to extract embedded features from several kinds of data and by assuming some mathematical assumptions \citep[e.g.,][]{Huang98}. Obviously, a suitable and well-built data analysis method should require to minimize mathematical assumptions and numerical artifacts, trying to maximize its adaptivity to the data under investigation \citep[e.g.,][]{Huang98}. A suitable method with the above characteristics is the well-known and well-established Hilbert-Huang Transform (HHT), firstly introduced by \citet{Huang98} as an adaptive and a posteriori data analysis procedure, mainly based on two different steps: a decomposition method, known as Empirical Mode Decomposition (EMD), and a statistical spectral method, e.g., the HSA \citep[e.g.,][]{Huang98}. Being $B_\mu(t)$ the $\mu$-th component of the interplanetary magnetic field, by means of the EMD and HSA we can write
\begin{equation}
    B_\mu(t) = \sum_{k=1}^N \mathcal{A}_{\mu, k}(t) \cos \left[ \Phi_{\mu, k}(t) \right] + \mathcal{R}_\mu(t), 
\end{equation}
being $\mathcal{C}_{\mu, k}(t) = \mathcal{A}_{\mu, k}(t) \cos \left[ \Phi_{\mu, k}(t) \right]$ the $k$-th empirical mode, $\mathcal{A}_{\mu, k}(t)$ and $\Phi_{\mu, k}(t)$ its instantaneous amplitude and phase, respectively, and $\mathcal{R}_\mu(t)$ the residue of the decomposition, e.g., a non-oscillating function \citep[e.g.,][]{Huang98}. More details about the HHT can be found in Appendix \ref{app:HHT}.

Although the HHT is surely interesting for investigating the multiscale behavior of physical systems, a distinguishing attribute is its suitability for investigating spectral and scaling features from a statistical point of view \citep[e.g.,][]{Huang11}. This can be done by defining the generalized marginal Hilbert power spectral density (gPSD) as
\begin{equation}
    \mathcal{S}_q(f) = \int_0^T \frac{\mathcal{H}_q(t',f)}{f} dt',
\end{equation}
being $T$ the time length and $\mathcal{H}_q(t',f)$ the generalized Hilbert-Huang spectrum accounting for the $q-$order amplitude distribution over the time-frequency plane \citep[cfr. Appendix \ref{app:HHT}, and][]{Huang11}. 
The scaling behavior of $\mathcal{S}_q(f)$ can be characterized by means of scaling exponents $\beta_q$ as
\begin{equation}
    \mathcal{S}_q(f) \sim f^{-\beta_q}, 
\end{equation}
being $\beta_q$ related to the scaling exponents $\zeta_q$ of the generalized structure functions $S_q(\tau) = |{B_\mu}(t+\tau) - {B_\mu}(t)|^q \sim \tau^{\zeta_q}$ as $\beta_q = \zeta_q+1$  \citep[e.g,][]{Huang11,Carbone18}.
However, due to its local nature, $\mathcal{S}_q(f)$ allows to determine scaling properties by reducing the effect of the noise, large-scale structures and inhomogeneities, and sampling effects \citep[e.g.,][]{Huang11}. 

\section{Results \& discussion} \label{sec:results}

\begin{figure}[ht!]
\plotone{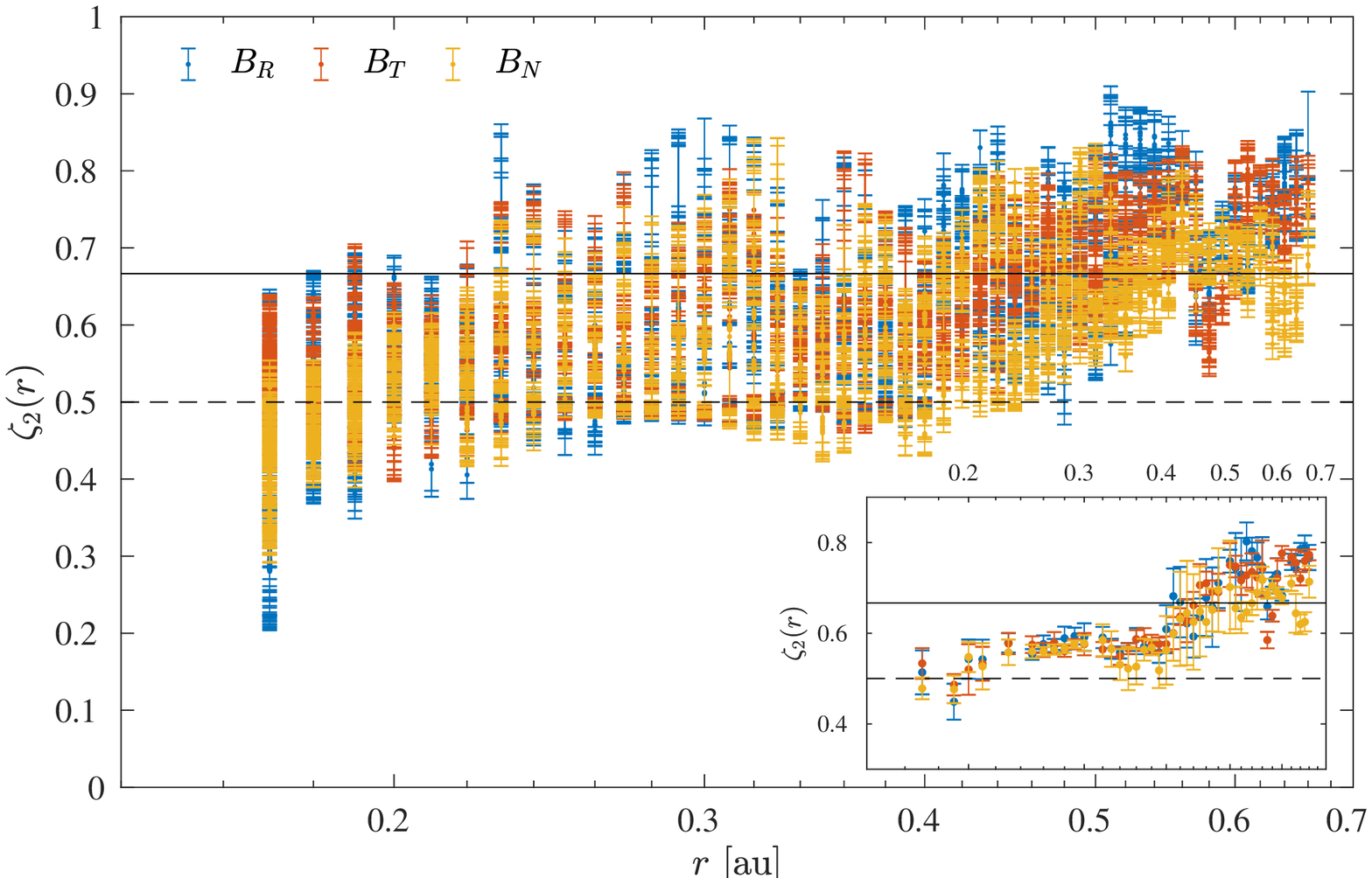}
\caption{The behavior of the scaling exponents $\zeta_2$ for each magnetic field component at different heliocentric distances $r$, together with the 95\% confidence level. The blue, orange, and yellow symbols refer to the radial $B_R$, tangential $B_T$, and normal $B_N$ components, respectively. The continuous and dashed black lines are used as a reference to $2/3$ \citep{Kolmogorov41} and $1/2$ \citep{Iroshnikov65,Kraichnan65} theoretical values, respectively. The inset show running averages at different heliocentric distances with a step $\Delta r = 0.01$ au (error bars are evaluated as the standard deviations). \label{fig2}}
\end{figure}

It has been widely shown that solar wind magnetic field fluctuations are characterized by a scaling law behavior in a wide range of frequencies, supporting the existence of an inertial regime where energy is transferred through an inviscid mechanism to higher frequencies \citep[e.g., to smaller scales,][]{Kolmogorov41,Iroshnikov65,Kraichnan65,Bruno13}. As recently pointed out by \citet{Chen20} spectral exponents move from $\alpha_B \sim$ -3/2 to $\alpha_B \sim$ -5/3 when passing from $r \sim$ 0.17 au to $r \sim$ 0.6 au, thus supporting the existence of a different energy transfer across scales. By means of the HHT we are able to investigate the behavior of scaling exponents $\zeta_q = \beta_q - 1$ of magnetic field components at different heliocentric distances by evaluating them for overlapping windows, at 1-hr steps, of length 1 day. Figure \ref{fig2} show the behavior of $\zeta(2)$ as a function of the heliocentric distance, together with the 95\% confidence level.

\begin{figure}[ht!]
\epsscale{0.85}
\plotone{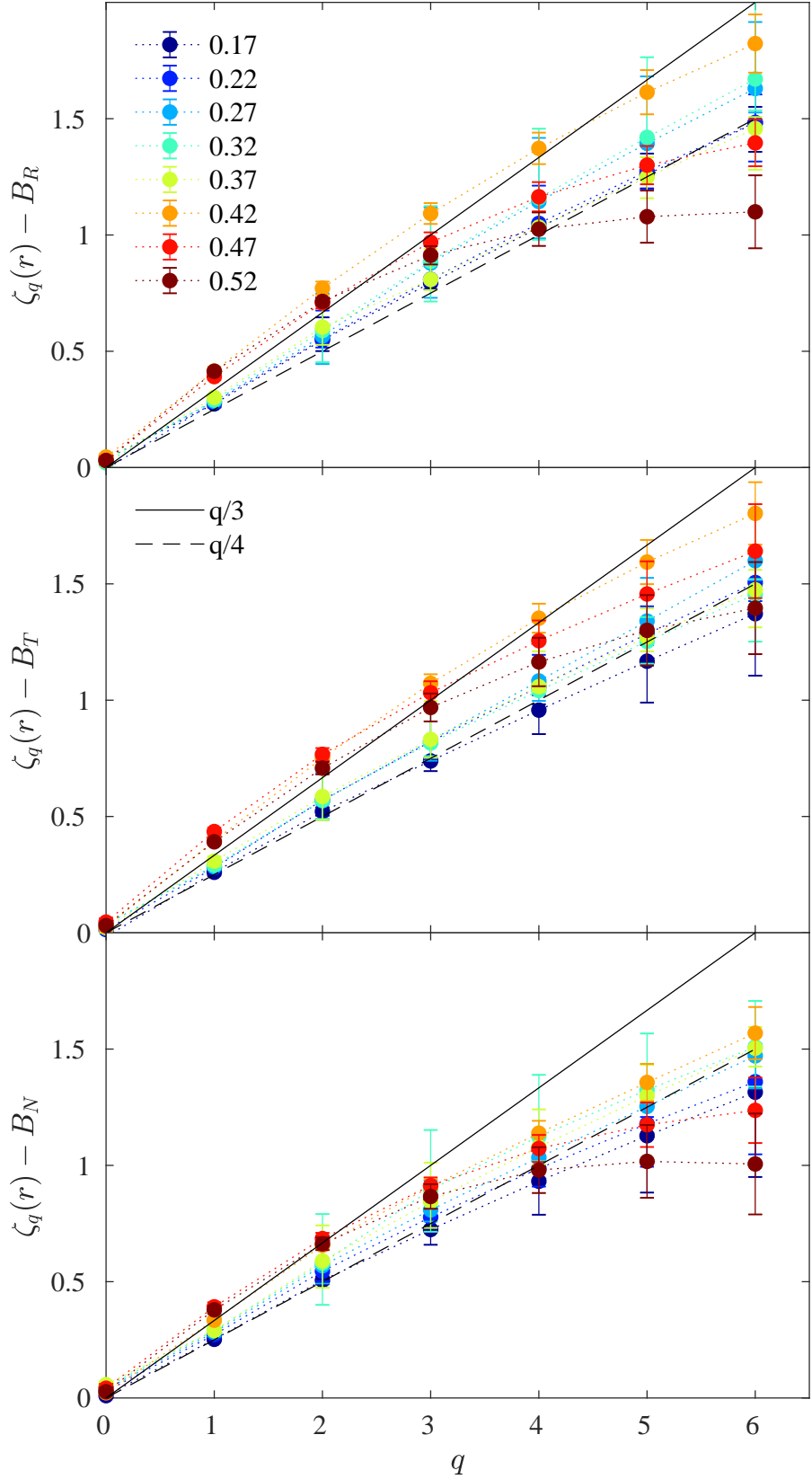}
\caption{The behavior of the scaling exponents $\zeta_q$ for each magnetic field component at different heliocentric distances $r$. The different colors correspond to different distances $r$ as reported in the legend. The continuous and dashed black lines are used as a reference to $q/3$ \citep{Kolmogorov41} and $q/4$ \citep{Iroshnikov65,Kraichnan65} theoretical scalings, respectively. Error bars show the 95\% confidence level. \label{fig3}}
\end{figure}

Results clearly show a difference between the scaling exponents $\zeta_2$ for distance below 0.4 au with respect to those evaluated at larger distances (i.e., larger than 0.4 au). This difference suggests that magnetic field fluctuations follows a $f^{-3/2}$ scaling closer to the Sun, being $\zeta_2 \simeq 1/2$, while a steeper scaling is found at larger distances ($\zeta_2 \simeq 2/3$ for $r > 0.4$ au). These findings are consistent with those reported by \citet{Denskat82} and \citet{Tu90} using Helios data, and more recently by \citet{Chen20} using PSP data. The lower $\zeta_2$ observed near the Sun could be related to a more steady-state nature of the inertial range, due to the large number of nonlinear times \citep{Matthaeus82}. Conversely, the larger values of $\zeta_2$ at $r > 0.4$ au can be related to a reduced value of the normalized cross-helicity as $r$ increases as well as to the role of intermittency \citep{Bruno13}. Both findings are also well in agreement with predictions made by numerical simulations of Alfv\'enic turbulence in homogeneous plasmas \citep{Boldyrev06,Lithwick07,Perez09,Chandran15,Mallet17}, suggesting that the inertial range processes vary from purely nonlinear interacting components to less organized fluctuations \citep{Velli89,Bruno13}. The transition from $\zeta_2 \sim 2/3$ to $\zeta_2 \sim 1/2$ as $r$ decreases gradually occurs and can be easily interpreted in the general framework of far-from-equilibrium complex systems as the evidence of a sort of dynamical phase transition which is consistent with the observed decreasing trend of positive correlation and the increasing of the outer scale with $r$ \citep{Chen20}. However, it is not sufficient to consider only one statistical moment of the probability distribution function to fully characterize solar wind turbulence. Indeed, since the pioneering work by \citet{Kolmogorov41} we know that turbulence is a phenomenon characterized by a hierarchy of scales whose statistics are scale-invariant \citep[e.g.,][]{Kolmogorov41,Iroshnikov65,Kraichnan65,Frisch95,Alberti19a}. The statistical scale-invariance implies that the scaling of field increments should occur with a unique scaling exponent, thus implying that the statistical moments of the field increments should scale as $S_q(\tau) \sim \tau^{q/D}$, being $D=3$ for fluid turbulence \citep[e.g.,][]{Kolmogorov41,Frisch95} and $D=4$ for plasma turbulence \citep[e.g.,][]{Iroshnikov65,Kraichnan65,Bruno13}. Nevertheless, there is considerable evidence that turbulent flows deviate from this behavior, being the scaling exponents a nonlinear function of the order $q$ \citep[e.g.,][]{Carbone95}, which point out an "anomalous" scaling  process and proves the appearence of intermittency  \citep[e.g.,][]{Frisch95,Bruno13}. For low orders the discrepancy with the linear behavior is very small, thus explaining why the Kolmogorov spectrum is usually observed in turbulence (e.g., $S_2(\tau) \sim \tau^{2/3} \to \mathcal{S}_2(f) \sim f^{-5/3}$). However, for high order statistics the difference is significant, and the breakdown of the statistical self-similarity is clear, thus questioning, in the modern theory of turbulence, what is really universal in the inertial range \citep[e.g.,][]{Alberti19b}. Thus, for a proper characterization we investigate the behavior of scaling exponents $\zeta_q$, $q \in [0,6]$, as derived from the generalized Hilbert PSDs $\mathcal{S}_q(f)$ (cfr. Section \ref{sec:methods}), at different heliocentric distances as shown in Figure \ref{fig3}.

\begin{figure}[ht!]
\plotone{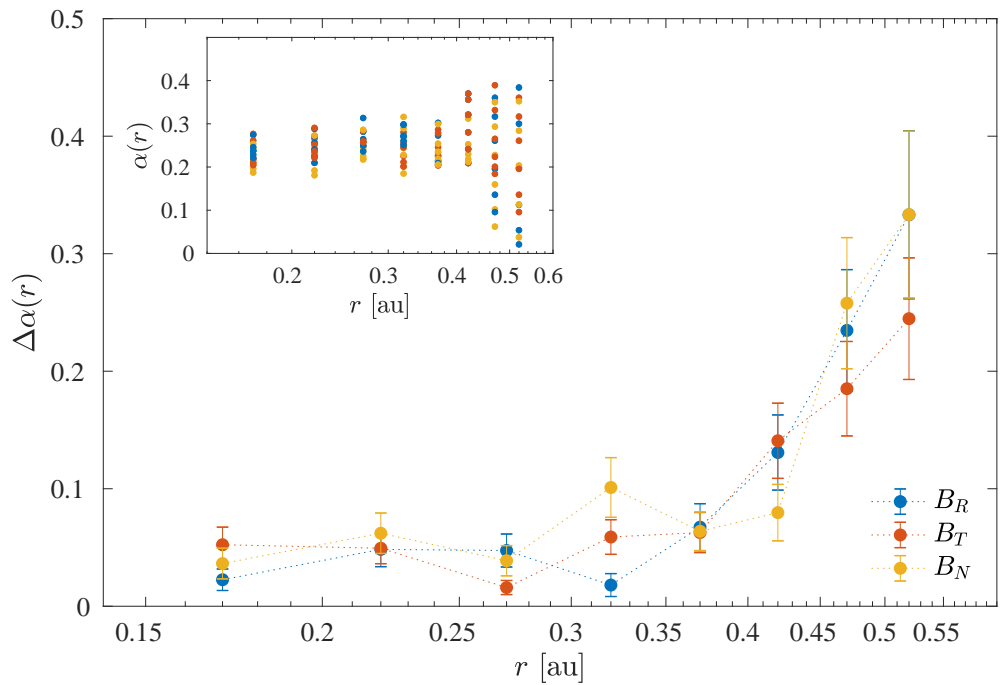}
\caption{The behavior of the multifractal width $\Delta \alpha(r)$ for each magnetic field component at different heliocentric distances $r$ as in Fig. \ref{fig3}. The blue, orange, and yellow symbols refer to the radial $B_R$, tangential $B_T$, and normal $B_N$ components, respectively. Error bars show the 95\% confidence level. The inset shows the behavior of $\alpha(r)$ at different distances $r$. \label{fig4}}
\end{figure}

Firstly, a clear difference emerges from the scaling behavior for $r <$ 0.4 au and for $r >$ 0.4 au: the former is linear with $q$, while the latter shows the typical convex nonlinear shape with $q$. The surprisingly behavior of scaling exponents near the Sun, suggesting a monofractal nature of field fluctuations within the inertial range, supports the assumptions of global statistical self-similar scale-invariance. Conversely, these assumptions break at 0.4 au, where the nonlinear convex behavior of scaling exponents, suggest a multifractal behavior of magnetic field fluctuations \citep[e.g.,][]{Bruno13,Alberti19a}. This transition could be related to physical processes suppressing the scaling properties of the energy transfer rate close to the Sun, being consistent with the emergence of intermittency in solar wind turbulence for $r > $ 0.4 au, also offering a novel scenario for the radial evolution of solar wind fractal nature for which, according to our knowledge, no exploration has been reported before in literature where only spectral features of field fluctuations were investigated at different locations \citep[e.g.,][]{Bavassano82,Denskat82,Grappin90,Marsch90,Tu90,Bruno13,Marsch18,Chen20}. The results suggest that, since the intrinsic nature of magnetic field fluctuations within the inertial range moves from monofractal to multifractal, then there should be a bifurcation parameter describing the observed changes into the scaling properties, opening a new perspective in the framework of dynamical systems \citep[e.g.,][]{Alberti19b}. The bifurcation parameter could be related to some plasma features as for example the $\beta$ parameter, the magnetic compressibility, the expansion/correlation time of fluctuations within the inertial range, the slow-/Alfv\'enic-mode variability within the heliosphere, the outward propagating Alfv\'enic fluctuations (predominantly originating from the Sun but undergoing a dynamical evolution due to nonlinear and velocity-shear), localized phenomena giving rise to intermittency, local changes in the cross-helicity, and so on \citep{Denskat82,Bavassano82,Matthaeus82,Tu90,Marsch90,Grappin90,Carbone95,Marsch18,Chen20}. Thus, the scaling exponents are not only a function of the statistical order $q$ but they also depend on the radial distance $r$ (i.e., $\zeta_q(r)$) which is the reflection of both global evolving and local dynamical processes. As also previously reported for spectral exponents, related to our findings by means of $\zeta_2$, at different heliocentric distances \citep[e.g.,][]{Denskat82,Marsch90,Tu90,Chen20}, there seems to be a change as the Sun is approached, rather suddenly inside 0.4 au \citep{Denskat82,Chen20}. Our findings not only strongly agree with seminal works when $q=2$ is considered \citep[e.g.,][]{Denskat82,Marsch90,Tu90,Chen20} but also allow, for the first time, to monitor the evolution of the scaling properties at different locations for high-order statistics, showing that the solar wind nature moves from monofractal to multifractal near 0.4 au. This change can be directly observed by looking at the behavior of singularities on the topology of solar wind magnetic field by means of the singularity strengths $\alpha(r) = \frac{d\zeta^{\mu}_q(r)}{dq}$ as usual in the multifractal approach \citep{Frisch95,Bruno13,Alberti19a}. In this way we can also provide a sort of multifractal measure $\Delta \alpha(r) = \max\{{\alpha(r)}\} - \min\{{\alpha(r)}\}$ (although we can only access the left part of the usual singularity spectrum $f(\alpha)$ since $q \ge 0$), thus allowing us to investigate the role of intermittency in changing the topology of the magnetic field. Fig. \ref{fig4} reports the behavior of the multifractal width $\Delta \alpha(r)$ for each magnetic field component at different heliocentric distances $r$ as in Fig. \ref{fig3}, while the inset show the behavior of singularity strengths $\alpha(r)$ at different distances $r$. 

We clearly observe a breakdown of the multifractal width $\Delta \alpha(r)$, moving from values closer to zero up to larger values $\Delta \alpha(r > 0.4) > 0.2$, thus suggesting the emergence of singularities as $r$ increases. This is confirmed by looking at the inset of Fig. \ref{fig4} in which is easy to detect a spread in singularity strengths $\alpha(r)$ as $r$ increases, with the transition observed near $r \sim 0.4$ au. 

\section{Conclusions} \label{sec:conclusions}

In this manuscript we dealt with the characterization of scaling features of magnetic field components as measured by PSP at different locations. We showed that the inertial range dynamics moves from a monofractal behaviour and a power spectrum scaling $f^{-3/2}$, at $r <$ 0.4 au, to a multifractal one and a power spectrum scaling $f^{-5/3}$, at $r >$ 0.4 au. This means that there is a transition region in which intermittency emerges, and the scaling properties of the inertial range are changed. Moreover, this also suggests that the solar wind magnetic field, in the early stages of its leaving the solar corona, show statistical self-similarity, while a breakdown of the statistical self-similarity for high-order statistics is found at a distance larger than 0.4 au from the Sun. In fact, we observed a roughly abrupt transition of the multifractal width $\Delta \alpha(r)$, moving from values closer to zero up to larger values $\Delta \alpha(r > 0.4) > 0.2$, thus suggesting that wider singularities are found at $r > 0.4$, also confirmed by the spread in singularity strengths $\alpha(r)$ as $r$ increases, with a transition observed near $r \sim 0.4$ au. Our results suggest that a dynamical phase transition occurs around 0.4 au and allow, for the first time, to characterize high-order statistics and the role of the intermittency in solar wind turbulence, suggesting that scaling exponents are not only a function of the statistical order $q$ but they also depend on the radial distance $r$ from the Sun, e.g., $\zeta_q(r)$, moving from a linear to a nonlinear convex behavior as $r$ increases. 

The observed transition could be related to something that suppresses the scaling properties of the energy transfer rate through the inertial range and the phase-coherency across the cascade for fluctuations close to the Sun. Roughly speaking, when the magnetic field is strong enough, since the scaling of the power spectra for inwards/outwards fluctuations are the same \citep{Chen20}, the usual Iroshnikov-Kraichnan model suggests that fluctuations should scales as $\langle (\Delta b)^q \rangle \sim c_A \langle \epsilon_\ell^{q/4} \rangle \ell^{q/4}$, instead of the usual Kolmogorov scaling $\langle (\Delta b)^q \rangle \sim \langle \epsilon_\ell^{q/3} \rangle \ell^{q/3}$ \citep{Bruno13}. In both cases, anomalous scaling laws $\zeta_q = hq + \mu(hq)$, being $h$ either $h=1/3$ or $h=1/4$, are recovered through the fluctuations of the energy transfer rate being $\langle \epsilon_\ell^q \rangle \sim \ell^{\mu(q)}$. The combined effect of the strong Alfv\'enicity and the reduced compressibility observed close to the Sun \citep{Chen20} should for example suppress the scaling behavior of the energy transfer rate, thus making $\langle \epsilon_\ell^{q/4} \rangle \sim const.$ for $r < 0.4$ au, while leaving $\langle \epsilon_\ell^{q/3} \rangle \sim \ell^{\mu(q/3)}$ far from the Sun, thus providing an explanation for our observations.

These considerations can be described in a general framework of far-from-equilibrium complex systems as the evidence of a dynamical phase transition for the fractal nature of solar wind magnetic field fluctuations at different heliocentric distances $r$. Indeed, the observed change from a monofractal to a multifractal nature suggest that there exists perhaps a bifurcation parameter which needs to be related to  plasma or wind parameters as the $\beta$ parameter, the magnetic compressibility, the expansion/correlation time of fluctuations within the inertial range, the slow-/Alfv\'enic-mode variability, the outward/inward propagating Alfv\'enic fluctuations, the localized emergence of velocity-shear and/or local changes in the cross-helicity, and so on \citep{Denskat82,Bavassano82,Matthaeus82,Tu90,Marsch90,Grappin90,Carbone95,Marsch18,Chen20}. In a simple conceptual model, being defined $\zeta^{\mu}_q(r)$ the scaling exponents of the magnetic field component $B_\mu(r)$ measured at the heliocentric distance $r$, can be written as 
\begin{equation}
    \zeta^{\mu}_q(r) = \sigma_\mu(r) \left( q + f(q, r) \right),
\end{equation}
being $\sigma_\mu(r)$ the bifurcation parameter and $f(q,r)$ a smooth nonlinear convex function of $q$ \citep[e.g.][]{Meneveau87,Carbone93,Bruno13}, slightly changing with $r$ as a sort of sigmoid function. This also simply traduces into a general radial evolution of singularity strengths as
\begin{equation}
    \alpha^{\mu}(r) = \alpha_0(r) + g(r, q),
\end{equation}
thus interpreting the inset of Fig. \ref{fig4} as a sort of bifurcation diagram resembling that derived in the case of saddle-node bifurcation, which can be also used for multifractal modeling purposes. 

The bifurcation parameter $\sigma_\mu(r)$ should depends, perhaps in a complex way, on the magnetic field intensity, the Alfv\'enicity of fluctuations and the presence of compressibility by slow-modes. Future orbits of PSP at smaller $r$ with hopefully a better temporal coverage of plasma parameters could allow to distinguish between the various possibilities. 

We think this study offers new perspectives for describing the fractal properties of solar wind and to correctly characterize turbulence and intermittency in space plasmas at different locations. Moreover, in our opinion the results can be particularly useful for building up novel multifractal cascade models, mostly starting from seminal works \citep[e.g.,][]{Meneveau87,Carbone93}, for providing and testing new phenomenological models of the MHD turbulence \citep[e.g.,][]{Lithwick07}, for considering the role of intermittency in modifying scaling features and scale-dependent behaviors \citep[e.g.,][]{Mallet17}, as well as to characterize the role of the large-scale forcing and decaying mechanisms on the inertial range cascade \citep[e.g.,][]{Chen20}. Further investigation will be devoted on the characterization the dynamical bifurcation occurring near $r = r_c \sim 0.4$ au in terms of a simple dynamical system admitting a saddle-node bifurcation as one or more control parameters are varied, although also different kind of bifurcations could be investigated (e.g., the super-critical pitchfork bifurcation) as well as its modeling in terms of (stochastic) Langevin systems or low-order discrete dynamical systems \citep{Alberti19b}.

\acknowledgments

The data used in this study are available at the NASA Space Physics Data Facility (SPDF), \url{https://spdf.gsfc.nasa.gov/index.html}. The FIELDS experiment on the Parker Solar Probe spacecraft was designed and developed under NASA contract NNN06AA01C. We acknowledge the contributions of the FIELDS team to the Parker Solar Probe mission. ML, GC, MFM, and VC thank the financial support by Italian MIUR-PRIN grant 2017APKP7T on Circumterrestrial Environment: Impact of Sun-Earth Interaction.

\appendix 
\section{The Hilbert-Huang Transform (HHT)} \label{app:HHT}

The first step of the Hilbert-Huang Transform (HHT), e.g., the Empirical Mode Decomposition (EMD), allows us to derive from the original signal ${B_\mu}(t)$ the set of empirical modes $\mathcal{C}_{\mu, k}(t)$. They are defined as functions having the same (or differing at most by one) number of extrema and zero crossings and a zero-average mean envelope and are obtained by means of the so-called sifting process based on the following steps:
\begin{enumerate}
    \item define the zero-mean signal ${B_\mu}_m(t) = {B_\mu}(t) - \langle{B_\mu}(t)\rangle$, being $\langle \dots \rangle$ the average value;
    \item find the local extrema of ${B_\mu}_m(t)$;
    \item find the upper $\mathcal{U}(t)$ and the lower $\mathcal{L}(t)$ envelopes by using a cubic spline;
    \item find the mean envelope $\rightarrow \mathcal{M}(t) = \frac{\mathcal{U}(t) + \mathcal{L}(t)}{2}$; 
    \item define $\mathcal{D}_\mu(t) = {B_\mu}_m(t) - \mathcal{M}(t)$;
    \item[6.] {\bf if $\mathcal{D}_\mu(t)$ is an empirical mode then}
    \begin{enumerate}
    \item[6.1] store $\mathcal{C}_{\mu, k}(t) = \mathcal{D}_\mu(t)$;
    \item[6.2] ${B_\mu}_m(t) \to {B_\mu}_m(t) = {B_\mu}_m(t) - \mathcal{D}_\mu(t)$; 
    \item[6.3] repeat steps 1.-5;
    \end{enumerate}
    \item[6.] {\bf if $\mathcal{D}_\mu(t)$ is not an empirical mode then}
    \begin{enumerate}
    \item[6.1] iterate steps 1.-5. until $\mathcal{D}_\mu(t)$ is an empirical mode;
    \item[6.2] store $\mathcal{C}_{\mu, k}(t) = \mathcal{D}_\mu(t)$;
    \item[6.3] ${B_\mu}_m(t) \to {B_\mu}_m(t) = {B_\mu}_m(t) - \mathcal{D}_\mu(t)$;
    \item[6.4] repeat steps 1.-5;
    \end{enumerate}
    \item[7.] stop the process when $\mathcal{R}_\mu(t) = \mathcal{D}_\mu(t)$ is a non-oscillating function or has only two extrema.
\end{enumerate}
Thus, a completely adaptive procedure is built, there are no assumptions and requirements on linearity and/or stationarity of ${B_\mu}(t)$, and the decomposition basis $\{\mathcal{C}_{\mu, k}(t)\}$ is a complete and orthogonal set, as for usual decomposition methods \citep[e.g., Fourier analysis or Wavelets,][]{Huang98}. 

The second step of the HHT is to investigate the amplitude and frequency modulation of each empirical mode by means of the so-called Hilbert Transform (HT) which is defined as
\begin{equation}
 \hat{\mathcal{C}}_{\mu, k}(t) = \frac{1}{\pi} \mathcal{P} \int_0^\infty \frac{\mathcal{C}_{\mu, k}(t')}{t - t'} dt',
\end{equation}
where $\mathcal{P}$ is the Cauchy principal value \citep[e.g.,][]{Huang98}. Then, by defining
\begin{equation}
 \mathcal{Z}_{\mu, k}(t) = \mathcal{C}_{\mu, k}(t) + i \, \hat{\mathcal{C}}_{\mu, k}(t) = \mathcal{A}_{\mu, k}(t) e^{i \, \Phi_{\mu, k}(t)},
\end{equation}
we can derive 
\begin{eqnarray}
 \mathcal{C}_{\mu, k}(t) &=& \Re\left\{{\mathcal{Z}_{\mu, k}}\right\} = \mathcal{A}_{\mu, k}(t) \cos\left[\Phi_{\mu, k}(t)\right], \\
 \mathcal{A}_{\mu, k}(t) &=& \sqrt{\mathcal{C}^2_{\mu, k}(t) + \hat{\mathcal{C}}^2_{\mu, k}(t)}, \\
 \Phi_{\mu, k}(t) &=& \tan^{-1} \left[ \frac{\hat{\mathcal{C}}_{\mu, k}(t)}{\mathcal{C}_{\mu, k}(t)} \right],
\end{eqnarray}
being $\mathcal{A}_{\mu, k}(t)$ and $\Phi_{\mu, k}(t)$ the instantaneous amplitude and phase of the $k-$th empirical mode, respectively, thus $\mathcal{C}_{\mu, k}(t)$ is modulated both in amplitude and phase \citep[e.g.,][]{Huang98}. Moreover, we can simply define the instantaneous frequency as $f_{\mu, k}(t) = \frac{1}{2 \pi}\frac{d \Phi_{\mu, k}(t)}{dt}$ and the mean timescale $\tau_{\mu, k} = \langle f_{\mu, k}^{-1}(t) \rangle_t$, with $\langle \dots \rangle_t$ identifying the time average. 

Despite the above interesting properties and features of the HHT, surely helpful for correctly identifying the multiscale behavior of physical systems, the HHT is particularly helpful for investigating spectral and scaling features from a statistical point of view \citep[e.g.,][]{Huang11}. Indeed, the combination of both EMD and HSA allows us to investigate how the energy content of a signal ${B_\mu}(t)$ evolves over different frequencies (i.e., at different timescales, allowing us a multiscale characterizations) and at different times \citep[e.g.,][]{Huang98}. This can be simply achieved by contouring in a time-frequency plane the square of instantaneous amplitudes of each empirical mode, thus defining the so-called Hilbert-Huang spectrum \citep{Huang98} 
\begin{equation}
    \mathcal{H}(t, f) = \mathcal{A}^2(t, f).
\end{equation}
The latter has a completely different meaning of energy spectra defined by means of other decomposition techniques, \citep[e.g., Fourier or Wavelet spectrograms,][]{Huang98}. Indeed, while for fixed scale decomposition methods the existence of energy at a frequency means that a component at that scale persisted through the whole time range, for the HHT it means that, in the whole time range, there is a higher likelihood for such a wave to have appeared locally, since frequency varies with time \citep[e.g.,][]{Huang98}. This is a direct consequence of the new concept of instantaneous frequency, thus implying that finding a frequency value $f^\star$ simply means that within the whole set of values of $f_{\mu, k}(t)$, $k=1, \dots, N$, there is a higher likelihood of finding the value $f^\star$ at the time $t^\star$ with a probability of $\mathcal{H}(t^\star, f^\star)$ \citep[e.g.,][]{Huang98}. Thus, the Hilbert-Huang spectrum acquires a statistical meaning, instead of having a more deterministic sense as for previous methods \citep[e.g.,][]{Huang98}. 

The concept can be rapidly expanded to all statistical moments of the instantaneous amplitudes probability distribution functions such that we can define \citep[e.g.,][]{Huang11}, for a given moment order $q \ge 0$,
\begin{equation}
    \mathcal{H}_q(t, f) = \mathcal{A}^q(t, f).
\end{equation}
As usual in statistics, by keeping fixed $q=2$ we account for the distribution of energy (e.g., the variance) at different frequencies $f$ and for any time $t$, and by integrating over time we account for the global energy distribution at different frequencies 
\begin{equation}
    \mathcal{H}_2(f) = \int_0^T \mathcal{H}_2(t', f) dt',
\end{equation}
known as Hilbert marginal spectrum \citep{Huang98}, directly related to the Fourier spectrum \citep[e.g.,][]{Huang11}. Finally, as firstly shown by \citet{Huang11} the generalized Hilbert-Huang spectra $H_q(t, f)$ can be powerfully used to investigate scaling law behavior of time series as well as to characterize fractal properties due to their analogy with standard structure function analysis \citep[e.g.,][]{Huang11,Consolini17,Carbone18}. Indeed, by integrating over time we can define 
\begin{equation}
    \mathcal{S}_q(f) = \int_0^T \frac{\mathcal{H}_q(t',f)}{f} dt'
\end{equation}
whose scaling behavior is equivalent to that of the generalized structure functions $S_q(\tau) = |{B_\mu}(t+\tau) - {B_\mu}(t)|^q$ \citep[e.g,][]{Huang11,Carbone18}. Indeed, while for structure functions the scaling behavior can be characterized by means of scaling exponents $\zeta_q$ as
\begin{equation}
    S_q(\tau) \sim \tau^{\zeta_q}, \label{eq:struct}
\end{equation}
for the HSA we have that 
\begin{equation}
    \mathcal{S}_q(f) \sim f^{-\beta_q}, \label{eq:sq}
\end{equation}
being \citep[e.g.,][]{Huang11}
\begin{equation}
    \beta_q = \zeta_q+1.
\end{equation}
Furthermore, if the exponents $\beta_q$ linearly behave with the order $q$ over a frequency range $f \in [f_1,f_2]$ then the process occurring within this range of frequencies is monofractal, while if $\beta_q$ is a nonlinear convex function of $q$ then it shows multifractal features \citep[e.g.,][]{Consolini17,Carbone18}.

\end{document}